# One-dimensional Active Contour Models for Raman Spectrum Baseline Correction

M. Hamed Mozaffari[*a] and Li-Lin Tay[a]

[a] *National Research Council Canada, Metrology Research Centre, Ottawa, ON, Canada.*

[*] *Corresponding author: email:* [mhamed.mozaffarimaaref@nrc-cnrc.gc.ca](mhamed.mozaffarimaaref@nrc-cnrc.gc.ca)

## 1 Abstract

Raman spectroscopy is a powerful and non-invasive method for analysis of chemicals and detection of unknown substances. However, Raman signal is so weak that background noise can distort the actual Raman signal. These baseline shifts that exist in the Raman spectrum might deteriorate analytical results. In this paper, a modified version of active contour models in one-dimensional space has been proposed for the baseline correction of Raman spectra. Our technique, inspired by principles of physics and heuristic optimization methods, iteratively deforms an initialized curve toward the desired baseline. The performance of the proposed algorithm was evaluated and compared with similar techniques using simulated Raman spectra. The results showed that the 1D active contour model outperforms many iterative baseline correction methods. The proposed algorithm was successfully applied to experimental Raman spectral data, and the results indicate that the baseline of Raman spectra can be automatically subtracted.

## 2 Introduction

Vibrational spectroscopy is a non-invasive technique for the analysis of chemical samples. This analytical technique (e.g., Raman spectroscopy) is used in a variety of applications, including security investigations (e.g., unknown substance detection [1-3]), material science (e.g., new semiconductor material for microelectronics [4]), biology and medicine (e.g., cancer cell identification, disease diagnosis, and drug discovery [5-7]), and space exploration (e.g., NASA rovers [8, 9]). In general, the Raman spectrum is a weak signal relative to noise. Although there are different techniques to enhanced Raman signal, including surface-enhanced Raman spectroscopy (SERS) [6, 10-13], still acquired Raman spectra contain undesirable features such as baselines.

State-of-the-art Machine learning (ML) techniques such as one-dimensional deep convolutional neural networks (1D CNNs) [14, 15] show that spectra analysis is feasible even with the existence of noise and background. However, these modern techniques' performance relies on the availability of a large database [16] of standard spectra comprising almost all possible background noise distributions. In contrast, available standard library databases are in limited numbers and often free of noise and baselines. Therefore, it is necessary to separate both background and noise from the signal itself prior to further spectral analysis.

Manual baseline correction by experts generally produces superior results, and for this reason, high accuracy manual baseline removal results are considered the gold standard [17]. However, manual baseline correction might not be feasible for circumstances such as real-time field applications and working with a large number of spectra. Therefore, it is vital for the field of spectral analysis to develop semi- or fully- automated baseline correction techniques [18, 19].

The active contour models [20] method is an implicit image segmentation [21, 22] technique that has been widely used in image processing and computer vision tasks. In this study, a modified version of active contour models in one-dimensional space is presented for fully automated and model-free baseline correction tasks in spectroscopy.

## 3 Methodology

### 3.1 One-Dimensional Active Contour Models

The basic idea of active contour models (sometimes called snakes) [20] is that an initialized curve interactively encompasses detected target object in an image. The curve has several control points (also called Snaxels or key-points). Control points are responsible for moving and deforming the curve. Internal and external energies are two constraints defined for each control point. Internal energies aim to keep key points separate enough and control curve deformation. External energies push or pull control points towards object features, including contours, edges, and lines [20]. In discrete form, the energy function of snakes [20] defines as follows:

$$E_{snake}^t = \sum_{i=1}^{N} \left( E_{int}^t(x_i^t) + E_{ext}^t(x_i^t) \right) \quad (1)$$

where $E_{int}^t(x_i^t)$ and $E_{ext}^t(x_i^t)$ are the internal and the external energies at the time t for an ith control point $x_i^t$ in 2D coordinates of x and y, respectively. N is the number of control points. The conventional energy functions [20] are defined as

$$E_{int}^t(x_i^t) = \alpha_i |x_i^t - x_{i-1}^t|^2 + \beta_i |x_{i-1}^t - 2x_i^t + x_{i+1}^t|^2 \quad (2)$$
$$E_{ext}^t(x_i^t) = -|\nabla f(x_i^t)| \quad (3)$$

where $f(x_i^t)$ is the original image pixel value at control point $x_i^t$ at time t. The constant coefficients of $\alpha_i$ and $\beta_i$ are defined to control the relative importance of each term on internal energy. The external energy is set to be the negative amount of the image gradient. Therefore, external energy is minimum at image locations with large gradient values (e.g., edges and lines). Through an energy minimization process, the location of control points converges to the desired object boundary. As the final step, a spline is fitted on the control points to delineate object contour.

There are many ways to minimize the energy of control points, e.g., gradient descent optimization [20], dynamic programming methods [23, 24], and heuristic techniques [25]. Also, variants of active contour models have been used for image segmentation in different applications, from 2D [20] to 3D [26] spaces, as standalone methods or a hybrid with others, such as deep learning techniques [27]. In this study, we proposed a modified version of active contour models for 1D applications. We utilized our method for the application of baseline correction in the Raman spectroscopy area.

### 3.2 Forces applied on control points

In classical mechanics, the principles of dynamics formulate what forces are applying to an object. Depends on these forces, objects tend to change their positions toward a new location with the lowest potential energy. Displacement specifications (i.e., direction, speed, acceleration, etc.)

depend on the net force applied to the object by other surrounding objects. To create a 1D active contour model, inspired by [28], we redefined control points' constraints. In this way, each control point is under the influence of three different forces. As we know, the first and second derivatives are large for positions with sharply inclined changes such as edges. Considering incremental unit time steps t, inclined force ($F_{inc}^t$) pull down or push up *i*th control point as follow:

$$F_{inc}^t(x_i^t) = \frac{\partial f(x_i^t)}{\partial x} + 0.5 \times \frac{\partial^2 f(x_i^t)}{\partial x^2} \qquad (4)$$

where $\frac{\partial f(x_i^t)}{\partial x}$ and $\frac{\partial^2 f(x_i^t)}{\partial x^2}$ are the first and the second derivatives of the 1D spectrum $f$ with respect to the *i*th control point position $x_i^t$ on the horizontal axis $x$. It is noteworthy to mention that Equ. (4) can be easily approximate with the constant acceleration kinematic formula in classical physics ($x = vt + 0.5 \times at^2$, ($t = 1, v_0 = 0, x_0 = 0$)) where $v = \frac{\partial x}{\partial t}$ and $a = \frac{\partial^2 x}{\partial t^2}$. We considered that control points are connected with linear springs. For this reason, control points experience two forces, one due to their elevation $F_{gra}^t(x_i^t)$ and the other due to the spring forces $F_{spr}^t(x_i^t)$. Equ. (5) and Equ. (6) show these two forces.

$$F_{gra}^t(x_i^t) = min\left\{\frac{-(f(x_i^t) - f(x_{i-1}^t))}{\sqrt[2]{(x_i^t - x_{i-1}^t)^2 + (f(x_i^t) - f(x_{i-1}^t))^2}}, 0\right\} + max\left\{\frac{(f(x_i^t) - f(x_{i+1}^t))}{\sqrt[2]{(x_i^t - x_{i+1}^t)^2 + (f(x_i^t) - f(x_{i+1}^t))^2}}, 0\right\} \qquad (5)$$

$$F_{spr}^t(x_i^t) = -k \times (x_i^t - x_i^{t-1}) \qquad (6)$$

Note that equation (6) is the famous Hooke's law in physics, and k is a constant factor characteristic of the spring. In general, the position of control points of 1D active contour model are updated from left to right by

$$x_i^{t+1} = x_i^t - \alpha F_{gra}^t - \gamma F_{inc}^t - F_{spr}^t \qquad (7)$$

The weighting factors in (7), $\alpha$ and $\gamma$, are defined to control the relative importance of each term. Figure 1 shows the updating procedure of control points using different forces in one iteration. Iteratively, control points converge to the optimum lowest position on the spectrum. The final 1D snake is a deformed curve that fits on the converged control points. In this study, we used 1D linear interpolation. Two end snaxels on the 1D snake could be fixed or free depends on the user's preferences and spectrum characteristics.

The area under the baseline curve is calculated for each iteration using Equ. (8). Depends on the user preference, the algorithm can report the converged curve at the maximum iteration range or the curve with the minimum area under the baseline curve.

$$C_T^t = \sum_{i=1}^{N} f(x_i^t) \qquad (8)$$

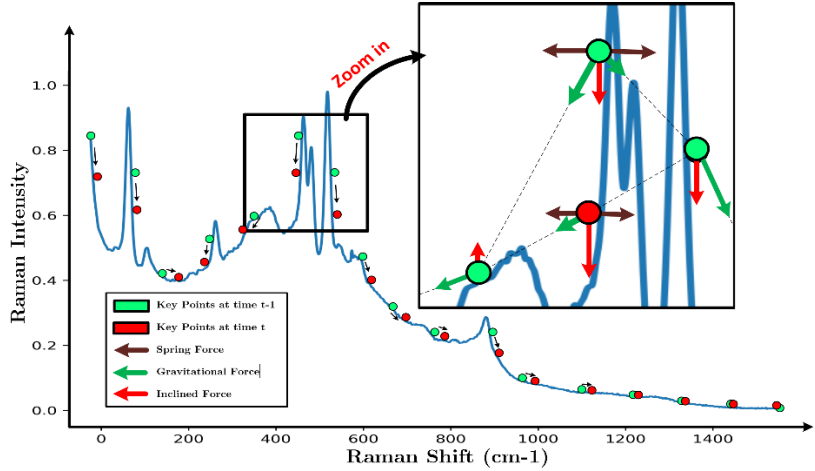

Figure 1 - A simple schematic of control point movements and forces are being applied on each iteration. Dashed lines are hypothetical lines for calculating forces. Note that the position of control points is updated in the x-axis direction.

## 4 Results and Discussion

We implemented our proposed idea in Python language running under Windows 10 on a PC with a 2.2GHz Intel CPU. An extensive hyperparameter tuning was performed to find the optimum parameters of 1D active contour models. We selected values of $\alpha$, $\gamma$, and $k$ as 0.3, 0.8, and 1.6, respectively. Note that the value of $k$ is associated with the number of initial snaxels, which we selected as 15. The number of iteration was set to 2500, and only the best area under the curve was reported. To evaluate our proposed technique, we used two databases of synthetic and real Raman spectra. The first database simulates, primarily, spectra obtained from vibrational spectroscopies such as Raman and infrared (IR). The real Raman spectra are SERS spectra of one chemical solution with different concentrations, acquired by a handheld Raman analyzer.

### 4.1 Synthetic Data

In theory, one acquired Raman spectrum can be modelled as the form $y = (s + b) * n_i + n_a$, where $s$ and $b$ are pure, and baseline signals convolve with instrumental blurring noise $n_i$. Ambient noise $n_a$ is also added to the signal. Therefore, the goal of baseline correction techniques is to estimate $s$. Solving the reverse equation $(s = (y - n_a) * n_i^{-1} - b)$ requires prior knowledge of $n_i^{-1}$, $n_a$, and $b$. Often in spectroscopy (e.g., for field operations), $n_i^{-1}$ is unattainable due to various unknown sources of noise. Also, the baseline is usually made from superimposing several different signals. Therefore, it is common to determine this knowledge by way of assumptions or estimations. For example, noise is often considered to be independent and with a normal distribution and with standard deviation $\sigma_n$ relative to the Raman spectrum intensity [19]. Similarly, the baseline is assumed to be broad, gradually increasing or decreasing, and with intensity below the Raman spectrum.

From the assumptions above, synthetic Raman spectra consisting of 1001 channels were created following the method mentioned in reference [19]. A sigmoid signal $b$ (made by cumulative normal function) was used as baseline. As can be seen from Figure 2a, the slope of the mid-point part of the baseline was selected to be one of the four values 0.0, 0.1, 0.3, or 1.0. Seven Lorentzian peaks (centers at channels 340, 455, 532, 584, 618, 641, and 656, respectively) were added to every baseline as pure Raman spectrum $s$. Full-width at half maximum (FWHM) of all Lorentzian peaks was set to 5.7 channels, except the peak at channel 340, which had an FWHM of 10 channels. Then the to simulate the instrumental noise of spectrometer $n_i$, pure Raman spectrum convolved with a 5-channel standard deviation normal distribution filter [19]. Figure 2b shows the pure synthetic Raman spectrum before superimposing on baselines. A 1001 zero-mean normal distribution was assumed for the ambient noise $n_a$. Then, standard deviation values for $n_a$ were determined using two definitions of the signal-to-baseline ratio (SBR) and the signal-to-noise ratios (SNR) as follow:

$$SBR = \frac{\max(s) - \min(s)}{\max(b) - \min(b)} \qquad (9)$$

$$SNR = \frac{\max(s) - \min(s)}{standard\ deviation\ (n_a)} \qquad (10)$$

By fixing values of SBR to 0.0, 0.05, 0.1, 1.0 and SNR to 2, 3, 5, 10, and 100, the standard deviation of noise signals can be calculated for each spectrum. Therefore, for every baseline $b$ with a given slope, five Raman spectra with Lorentzian peaks were generated. Then, five noise distributions were added to each of the generated spectra. In total, 100 synthetic Raman spectra ($4b \times 5(s * n_i) \times 5n_a$) were generated. Note that similar to the method of [19], we added a bias of one for baselines and assumed the $\min(b)$ as zero for Equ. (9).

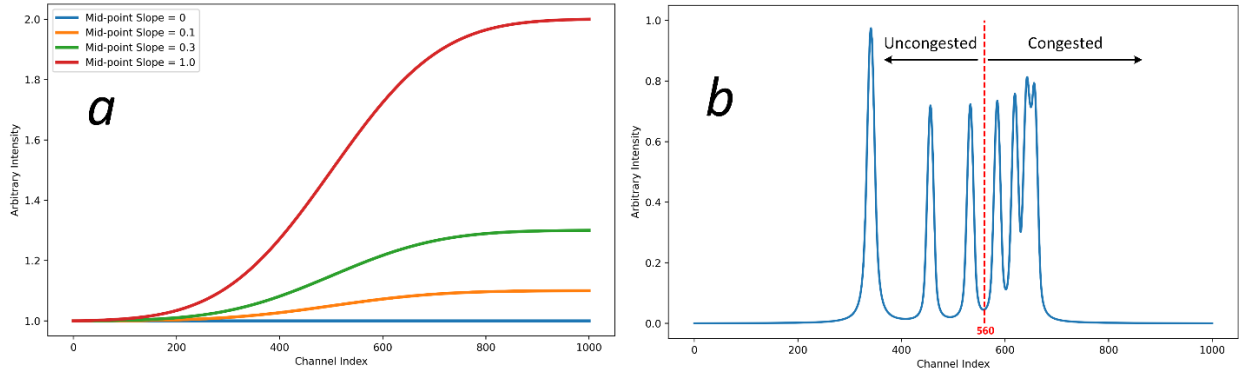

Figure 2 – Synthetic Raman data. (a) Baseline sigmoid functions used in standardized Raman spectra with different slope values of the linear part. (b) Simulated Raman spectrum consisting of seven Lorentzian peaks with two congested and uncongested regions.

Several baseline correction techniques have been investigated in [19] using bulk Figures-of-Merit (FOM) method. Likewise, we assessed the capability of our proposed method using the 100 generated synthetic Raman spectra. The correlation coefficient ($r$) was calculated between estimated and given baselines for uncongested, congested and entire regions of Raman spectra. Furthermore, for the entire region, chi-squared ($X^2$) was also calculated [19]. The first criterion presents the quantitative differences between actual baselines (denoted as ground truth (GT) baselines) and estimated ones, while the second one indicates the offsets between GTs and estimated baselines. Note that for the case of zero slope in GT baseline, the covariance matrix is undefined, therefore, correlation coefficient value (r) cannot be defined. Figure 3a shows bar graphs of correlation coefficients and chi-squared values between GT and estimated baselines. Also two examples of this experiment were shown in Figure 3b.

Figure 3a shows that the method produced relatively accurate baselines for all congested, uncongested, and complete regions when SBR values are small. However, it can be seen that the method tends to have more difficulties with spectra with SBRs 1.0 and 10.0. From [29], large values for $X^2$ in these cases might be low convergence rate of the method. Negative values for $r$ in these cases also related to the linear interpolation between snaxels indicate that 1D active contour models require further iterations. At the same time, low performance of the method on uncongested regions than congested ones might be an indication that some snaxels stuck in local minimas. Average values for $r$ and $X^2$ for 1D active contour model were 0.78 and 1106.9, respectively. From the global comparison results of reference [19], our method is better than almost all other iterative techniques except Artificial Neural Networks and signal removal methods. We determined optimum parameters of the 1D active contour method using synthetic spectra and grid search technique. Further investigation on several individual synthetic spectra that our proposed approach failed to achieve acceptable $r$ and $X^2$ revealed that by selecting accurate parameters and interpolation technique, it can estimate baseline accurately even for those failed cases.

### 4.2   Real Raman Data

Using a handheld spectrometer "ReporterR" from SciAps company, we acquired spectra of Benzenthiol solutions with two concentrations. The "ReporterR" was equipped with 70 mW, 785 nm excitation laser, a 2048 pixel TE cooled CCD array detector, and 1800 lines/mm grating. The spectrometer with the resolution of 12 cm$^{-1}$ was calibrated using the manufacturer-supplied polystyrene reference standard (more details can be found in [13, 14]). Figure 4 shows the results of applying our method to two spectra acquired with solution concentrations of 5$\mu M$ and 1$\mu M$. Parameters for this experiment were the same as the results of hyperparameter tuning values. The figure shows that the method performs better on the SERS spectrum of higher solution concentration than the lower one. Note that the location of two boundary snaxels was fixed.

## 5   Conclusion

This paper presented a baseline correction algorithm for Raman spectroscopy using a 1D modified active contour model (1D snake). Automatically initialized snaxels iteratively converge toward the desired Raman spectrum baseline. Then, an estimated baseline can be generated by interpolation between converged keypoints. Our method was evaluated on several synthetic and experimental SERS spectra. Experimental results showed that our approach is capable of estimating the baseline of Raman spectra automatically. The experimental results of the simulated spectra demonstrated that the 1D active contour model algorithm provided better performance for baseline estimation than many iterative techniques. The baseline correction results of the experimental Raman spectra also showed that our method could handle various types of backgrounds for real Raman spectra. The same parameters were employed throughout all of our experiments. We believe that the 1D snake technique can estimate baseline from any spectroscopic data type by selecting suitable snake parameters.

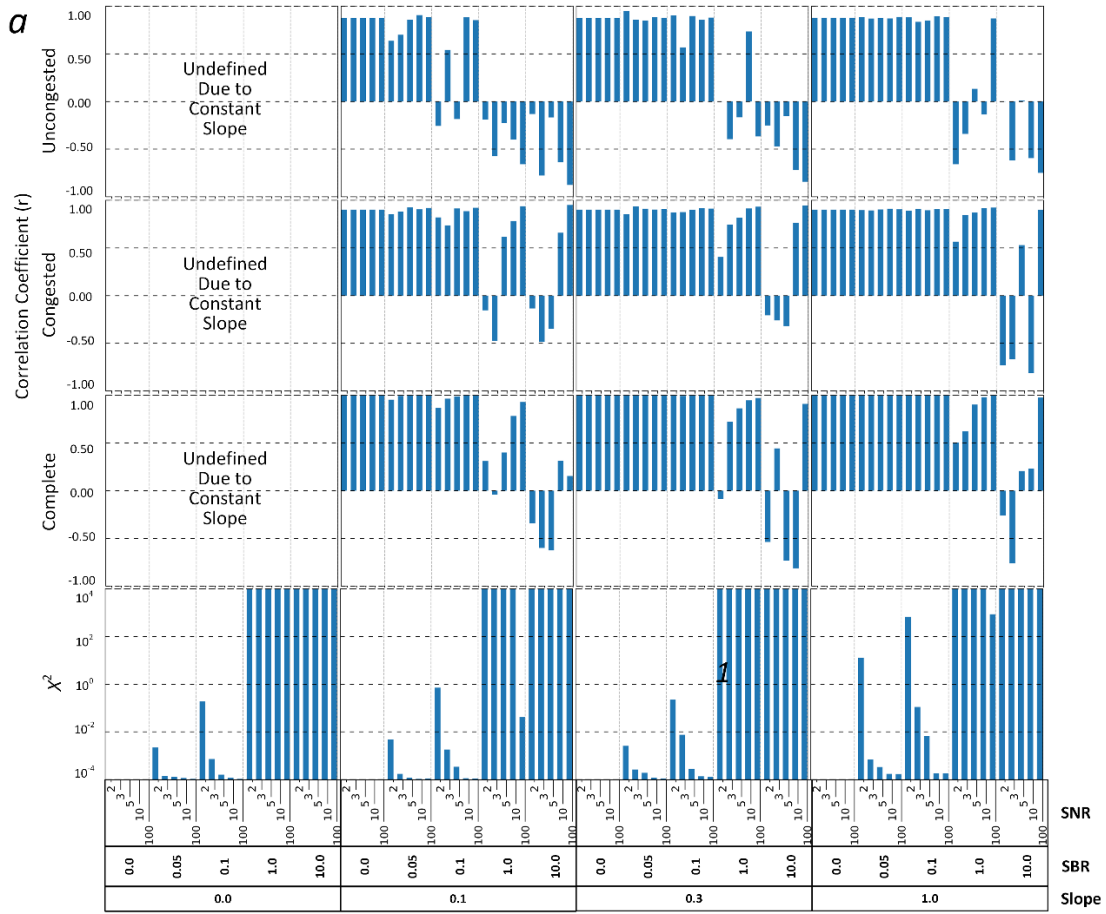

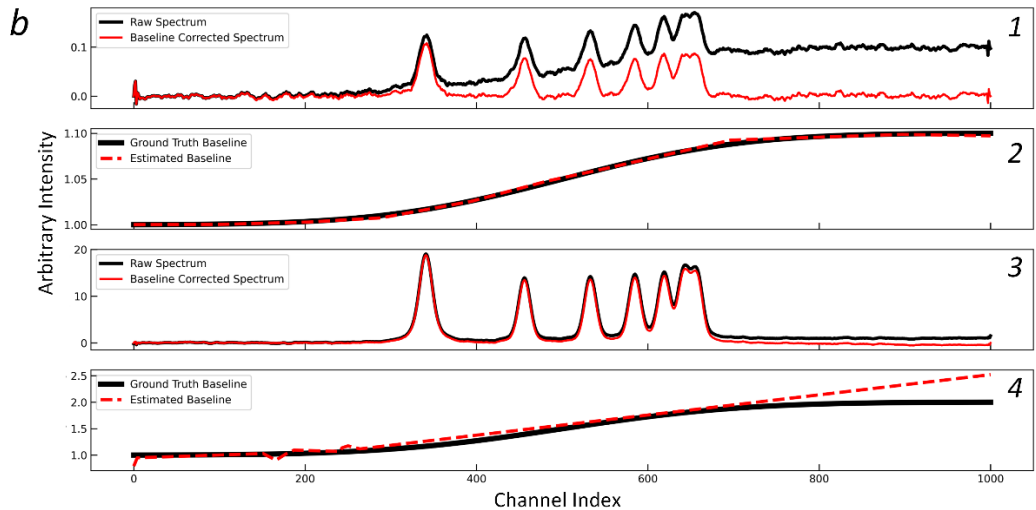

Figure 3 – Baseline correction using the 1D active contour model method. (a) The correlation coefficient between theoretical and estimated baselines for the uncongested, congested, and complete spectral regions of 100 synthetic spectra. Also, the chi-squared ($X^2$) was shown for the case of the complete spectral region. (b) (1) initial input spectrum with SNR = 10, SBR = 0.1, and baseline slope = 0.1 and its baseline corrected version using our method; (2) estimated baseline and ground truth sigmoid baseline with the same parameters as for (1); (3, 4) repeating the same scenario as for (1) and (2) with SNR = 100, SBR = 10, and baseline slope = 1.0.

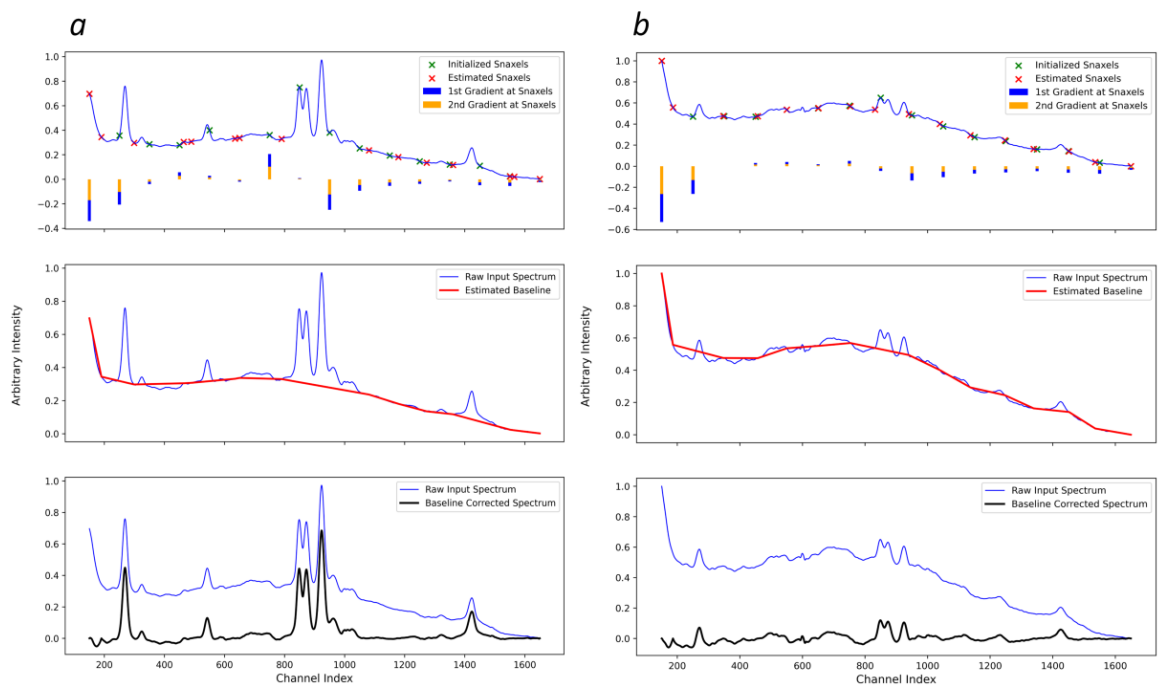

Figure 4 – Baseline removal using 1D active contour model for two raw Raman spectrum of Benzenthiol with concentrations of (a) 5$\mu M$ and (b) 1$\mu M$. Green and red color snaxels in the first row are initialized and estimated key points, respectively, where two blue and yellow bars indicate the value and direction of gradients in initialized snaxels. The second row's estimated baseline was generated from red snaxels in the first row using linear interpolation.